# An Ultra-High-Vacuum Rotating Sample Manipulator with Cryogenic Cooling

X. Y. Tee, A. Paré, A. P. Petrović, C. Panagopoulos[a]

*Division of Physics and Applied Physics, School of Physical and Mathematical Sciences, Nanyang Technological University,*
*21 Nanyang Link 637371, Singapore*

(a) Author to whom correspondence should be addressed: christos@ntu.edu.sg

Abstract – We report a homebuilt ultra-high-vacuum (UHV) rotating sample manipulator with cryogenic cooling. The sample holder is thermally anchored to a built-in cryogenic cold head through flexible copper beryllium strips, permitting continuous sample rotation. A similar contact mechanism is implemented for the electrical wiring to the sample holder for thermometry. The apparatus thus enables continuous sample rotation at regulated cryogenic temperatures in a UHV environment. We discuss the potential applications of this apparatus for cryogenic sputtering.

Fabrication and test processes often require samples or substrates to be rotated in a variable-temperature vacuum environment, while preserving electrical access to the sample for monitoring and/or analysis. Here we report a homebuilt UHV sample manipulator offering both continuous sample rotation and cryogenic cooling, with electrical connections to the sample holder. This instrument comprises rotating and non-rotating clusters of components. The rotating cluster consists of a central shaft extending from the mounting flange to the sample holder, with several other components attached. The rotating sample holder is thermally anchored to the non-rotating cryogenic cold head through flexible copper beryllium strips. Electrical connections between the rotating and non-rotating parts are made through a similar contact mechanism. Cryogenic techniques are implemented to reduce heat load to the sample holder, with radiation baffles and an enclosure shield used to block thermal radiation from entering the sample holder. The sample holder is deliberately positioned at a distance from the mounting flange to reduce the thermal conductance and hence heat transfer between the two ends. These efforts greatly reduce the undesirable heat load to the sample holder, enabling it to achieve a base temperature below 50 K when the apparatus is cooled by liquid helium at a flow rate of 1 liter per hour.

Figure 1 shows an overview of the UHV sample manipulator, which has a DN150CF mounting flange and a length of 1120mm as measured from the bottom of the flange. The mounting flange also contains four DN16CF flanges for electrical feedthroughs and a central DN16CF flange housing a mechanical rotary feedthrough. A thin-walled stainless steel shaft of O.D. 11mm is attached to the rotary feedthrough and extends 1105mm to the sample holder. A linear manipulator is placed between the rotary feedthrough and the central DN16CF flange to enable motion of the sample holder in the z-axis direction. Below the mounting flange, the structure of the manipulator can be conveniently divided into three functional units: namely the cold head, rotatable sample holder, and rotary contact mechanism for electrical wiring. Thin wall stainless steel tubes are used for the cryogen supply lines as well as mechanical support for the apparatus.

The cold head unit (Fig. 2) consists of several components all serving one purpose, namely the establishment of a stable and sufficiently low base temperature at the sample holder. Cooling power is provided by cryogenic liquids such as liquid helium or nitrogen, which flow into a coil reservoir through dedicated cryogen supply and return tubes welded to the mounting flange. The coil reservoir is made of OFHC copper tubing of O.D. 90mm and a length of 58mm. For improved cooling efficiency, it is brazed to the cold head which is a massive cylindrical OFHC copper block of diameter 90mm and height 58mm. The cold head has a clearance hole allowing through passage of the central rotating shaft. Flexible silver-plated copper beryllium strips are evenly mounted to the lower side of the cold head. Each strip has original dimensions (L x W x t) of 32mm x 6.25mm x 0.25mm. The strips are then bent into "L-shape" leaf springs of height 16 mm. The cold head is enclosed by a radiation shield made entirely of aluminium. Radiation baffles made of thin aluminium discs are evenly spaced along the manipulator body. The upper sides of these discs are mirror polished for enhanced reflectivity.

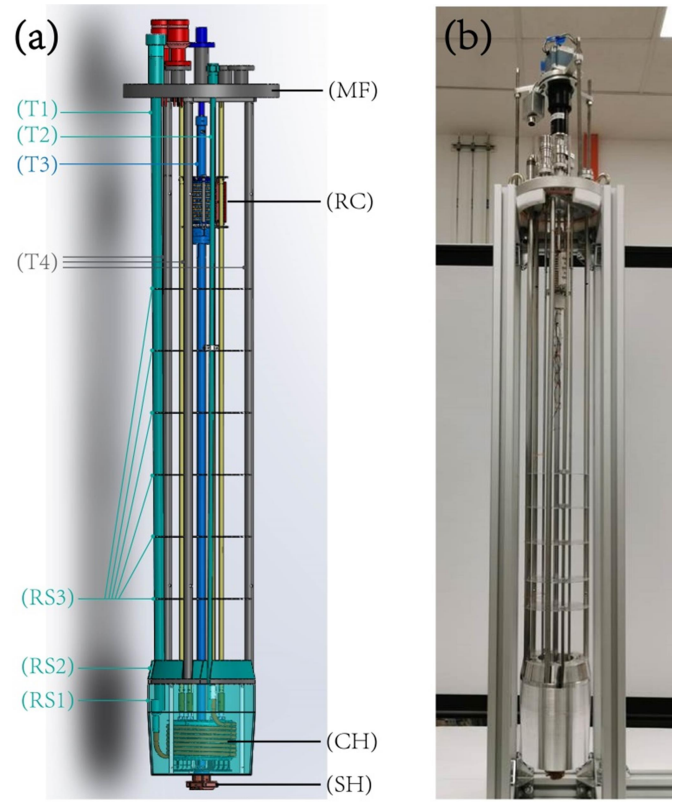

**Figure 1:** **(a)** The UHV sample manipulator and its components. MF is the mounting flange; RC is the rotary contact mechanism; CH is the cold head to which the cryogen reservoir coil is brazed and SH is the sample holder. T1 and T2 are the liquid helium transfer and return tubes, respectively; T3 is the central shaft which drives the rotation of the sample stage; T4 are three thin-walled stainless steel support tubes for the manipulator. RS1 and RS2 are the wall and lid of the radiation enclosure shield, respectively, and RS3 are radiation baffles which are evenly spaced above the radiation enclosure shield. **(b)** A photograph of the UHV sample manipulator.

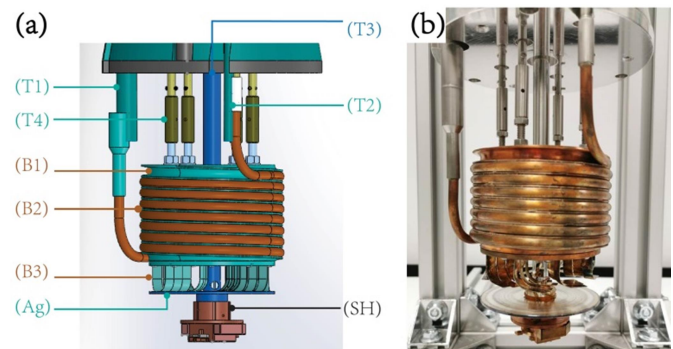

Figure 2: (a) The cold head unit and its components. T1 and T2 are the liquid helium supply and return tubes; T3 is the central rotation shaft; T4 are support tubes for the mechanical stabilization of the cold head. B1 is a cylindrical OFHC copper block with a clearance hole for through passage of the central rotation shaft; B2 is the coil reservoir for the liquid cryogen; B3 are flexible copper beryllium strips which create the thermal contacts to the sample holder (SH). Ag is the silver disc attached to the sample holder. (b) A photograph of the cold head unit.

The sample holder (Fig. 3) is attached to the central shaft for controlled continuous rotation. It is made of OFHC copper to ensure uniform temperature throughout the holder. A silicon diode thermometer and a cylindrical cartridge heater (25Ω, 6.4mm x 25.4mm) are mounted on the holder using UHV-compatible epoxy, thus facilitating precise temperature control. These electrical components are covered by copper beryllium foil shields. A sample of dimensions up to 10mm x 10mm can be transferred from a side-access load-lock chamber directly to the sample holder using a wobble stick. The sample holder is located just below the cold head. A silver disc of O.D. 80mm and thickness 1.5mm is mounted on the top side of the sample holder. The silver disc forms a rotatable surface in contact with the non-rotating silver-plated copper beryllium strips of the cold head. This establishes a strong thermal link which delivers substantial cooling power to the sample holder. When high sample temperatures are required, the silver disc can be moved *in-situ* using the external linear manipulator to reduce the pressure from (and eventually lose contact with) the copper beryllium strips. This cuts off cryogenic cooling and reduces the thermal mass associated with the sample holder, enabling efficient sample heating with the cartridge heater.

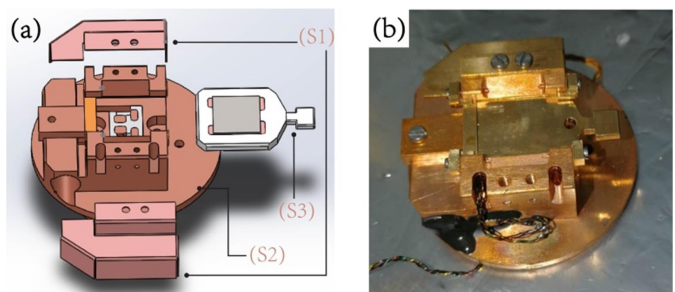

Figure 3: (a) The sample holder and its components. S1 are shield covers under which the thermometer and heater were mounted; S2 is the body of the sample holder; S3 is a sample loader upon which a 10mm x 10mm substrate is mounted for ease of sample transfer. The sample loader is transferred from a side-access load-lock chamber and inserted into the sample holder using a wobble stick. (b) A photograph of the sample holder.

The rotary contact mechanism (Fig. 4) is an interface between the incoming electrical wires from the mounting flange and the lower set of electrical wires running to the sample holder for the thermometer and the heater. The incoming wires are clamped to a connector plug. Here, each wire is connected to miniature copper beryllium strips which are electrically conducting. The lower set of wires is clamped to a connector receiver which connects the wires to a series of copper rings wound around the cylindrical body of the receiver. The receiver has a circular clearance hole for through passage of the central rotating shaft. However, these two components are mechanically locked together and hence rotate in phase. This is necessary to avoid twisting and eventually breaking the wires as the shaft rotates. The bodies of the connector plug and receivers are made of Teflon for electrical insulation. As the connector receiver rotates, the copper rings form a continuous running surface to the miniature copper beryllium strips on the plug. An electrical connection is established when the two are pressed into contact.

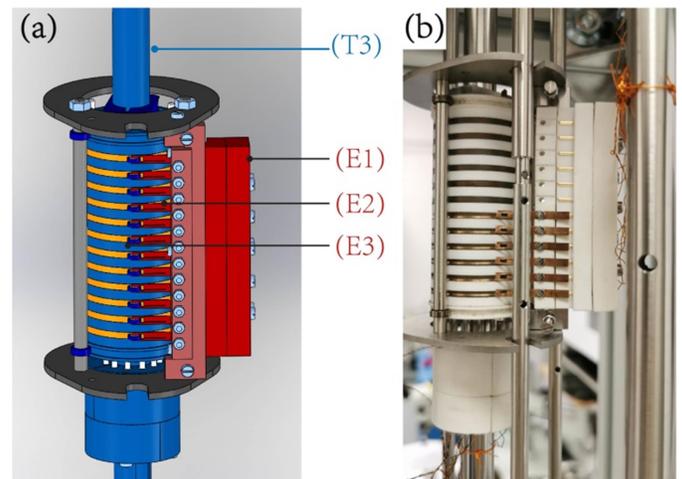

Figure 4: (a) The rotary contact mechanism and its components. T3 is the central rotation shaft. E1 is the insulation clamp made of Teflon; E2 are the miniature copper beryllium strips. E1 and E2 form the connector plug; E3 is the connector receiver with copper rings wound around its cylindrical body, which in turn is locked to the central rotation shaft. (b) A photograph of the rotary contact mechanism.

For tests, the manipulator was mounted on an UHV chamber. The base pressure of the chamber reaches $1 \times 10^{-8}$ mbar after an overnight pumpout, which confirms the UHV compatibility of the apparatus. The cryogenic cooling performance was tested with liquid helium, supplied at a flow rate of 1 liter per hour. This cooled the sample holder to a temperature of 50 K in 5.5 hours. A further 5-10K reduction in temperature can be achieved if necessary, either by cooling for a longer period or by increasing the LHe flow rate. Continuous sample stage rotation was driven by a stepper motor and its rotational speed was tested up to 30 rpm.

The sample manipulator was designed for but not limited to deposition systems. To provide context, we discuss its potential application for cryogenic sputtering. In the absence of impurities, the quality and properties of a deposited film can be controlled by the substrate temperature[1,2]. Elevated substrate temperatures are typically employed to improve the crystallinity of deposited films. A substrate at cryogenic temperature, on the other hand, quenches the impinging atoms to achieve smaller grain sizes and enhanced surface homogeneity[3,4]. Cryogenic sputtering under certain process conditions is able to preserve the crystalline structure[4] and epitaxial growth is possible[5,6].

Sputtering systems cooled by liquid cryogens have already been reported[6–8]. Cooling is typically achieved through the delivery of cryogenic liquid to a cold head upon which the substrate holder is directly mounted[7,8]. This provides effective cooling to the substrate holder. The base temperature is typically in the range of 40-70K, which is sufficiently low to achieve the beneficial effects of cryogenic sputtering[4,6–8]. A major drawback in these systems, however, is the lack of substrate rotation which is crucial for *confocal* deposition - a configuration used to deposit complex composite materials and multilayer heterostructures. The confocal configuration refers to a geometrical layout where multiple sputter targets are oriented at incident angles in the range of 15-30 degrees with respect to the substrate normal. Without substrate rotation, the sputter atoms land on the substrate at large incident angles, leading to undesirable non-uniform deposition across the substrate. The substrate rotation averages out the incident angles of the impinging atoms thus ensuring uniform thickness and composition for the deposited films[9]. Furthermore, the substrate rotation suppresses the so-called "shadow effect" which tend to develop microstructures for atoms impinging at large incident angles[10]. To date, there has been no report of confocal deposition at cryogenic temperatures."

There are relatively few reports for simultaneous provision of cryogenic cooling and electrical wiring to a rotating sample holder, but we are aware of two US patents relevant to this work. Josephs et al proposed a different design for a rotating liquid nitrogen cooled substrate holder[11]. However, it was not clear if the apparatus can be operated with liquid helium and whether it is UHV compatible. Also, their design does not incorporate a manipulator for the axial motion of the sample. Lubomirsky et al proposed an industrial solution for a rotating temperature-controlled substrate pedestal cooled by chilled water[12]. However, the proposed apparatus was not designed for cryogenic cooling. In particular, the constant cooling of the rotary fluid seal will cause vacuum leakage if cryogenic fluids are used.

In summary, our homebuilt sample manipulator presents a unique solution to integrate UHV compatibility, sample rotation, cryogenic cooling, and axial manipulation in one standalone apparatus. We discussed its impact to the field of cryogenic sputtering and note its potential for performing cryogenic growth of complex composites and multilayer heterostructures.

DATA AVAILABILITY

The data that support the findings of this study are available from the corresponding author upon reasonable request.

ACKNOWLEDGEMENTS

We acknowledge support from the National Research Foundation (NRF), NRF-Investigatorship (No. NRFNRFI2015-04), and Singapore MOE Academic Research Fund Tier 3 Grant MOE2018-T3-1-002. We are grateful to Ferrovac GmbH for technical assistance.